\documentclass{frontiersSCNS}

\usepackage{url,hyperref,lineno,microtype,subcaption}
\usepackage[onehalfspacing]{setspace}
\usepackage{url}
\usepackage{nicefrac}

\usepackage{xcolor} 
\hypersetup{                              
    colorlinks = true,                    
    citecolor = {blue},
    linkcolor = {green},
    urlcolor  = {purple},
           }

\graphicspath{{./}{./figs/}}

\def\keyFont{\fontsize{8}{11}\helveticabold }
\def\firstAuthorLast{S\'andor {et~al.}}
\def\Authors{
Bulcs\'u S\'andor\,$^{1,2,*}$, 
Michael Nowak\,$^{2}$, 
Tim Koglin\,$^{2}$,
Laura Martin\,$^{2}$ and 
Claudius Gros\,$^{2}$
}
\def\Address{
$^{1}$Department of Physics, 
Babes-Bolyai University, 
Cluj-Napoca, Romania\\
$^{2}$Institute for Theoretical Physics, 
Goethe University Frankfurt, 
Frankfurt am Main, Germany}
\def\corrAuthor{Bulcs\'u S\'andor}
\def\corrEmail{bulcsu.sandor[.]phys.ubbcluj.ro}

\begin{document}
\onecolumn
\firstpage{1}

\title[]
{Kick control: using the attracting states arising within the
sensorimotor loop of self-organized robots as motor primitives}

\author[\firstAuthorLast ]{\Authors}
\address{}
\correspondance{}
\extraAuth{}

\maketitle

\begin{abstract}
Self-organized robots may develop attracting
states within the sensorimotor loop, that is within
the phase space of neural activity, body and environmental
variables. Fixpoints, limit cycles and chaotic attractors
correspond in this setting to a non-moving robot, to
directed, and to irregular locomotion respectively. Short
higher-order control commands may hence be used to 
kick the system from one self-organized attractor 
robustly into the basin of attraction of a different 
attractor, a concept termed here as kick control. 
The individual sensorimotor states serve in this context 
as highly compliant motor primitives.

We study different implementations of kick control for the case of
simulated and real-world wheeled robots, for which
the dynamics of the distinct wheels is generated
independently by local feedback loops. The feedback 
loops are mediated by rate-encoding neurons 
disposing exclusively of propriosensoric inputs in
terms of projections of the actual rotational angle
of the wheel. The changes of the neural activity are
then transmitted into a rotational motion by a simulated
transmission rod akin to the transmission rods used for
steam locomotives.

We find that the self-organized attractor landscape 
may be morphed both by higher-level control signals, in
the spirit of kick control, and by interacting with the 
environment. Bumping against a wall destroys the limit 
cycle corresponding to forward motion, with the consequence 
that the dynamical variables are then attracted in phase space 
by the limit cycle corresponding to backward moving. The 
robot, which does not dispose of any distance or contact
sensors, hence reverses direction autonomously.

\section{}
\tiny
\keyFont{ \section{Keywords:} closed-loop robots, limit cycles, 
sensorimotor loop, self-organized locomotion, compliant robot,
robophysics} 
\end{abstract}

\section{Introduction}

The sensorimotor system is in general a product of evolution, 
development, learning and adaptation \citep{todorov2004optimality}.
One may examine alternatively whether self-organizing
principles \citep{prokopenko2009information} are capable 
to generate locomotion, in particular for the case of embodied 
\citep{ghazi2017morphological} and/or biologically inspired 
robots \citep{pfeifer2007self}. Self-organization may serve in 
this context to generate a palette of behavioral 
primitives \citep{tani2003self}, or, on a higher level,
to generate complex and playful behavior \citep{martius2013information}.

Attracting states in the sensorimotor loops corresponding to
regular and exploratory motion, that is respectively to limit 
cycles and chaotic attractors, can be generated following two 
complementary routes. Within the first approach, which is especially 
suited for settings involving a large number of degrees of
freedom \citep{kubisch2011using}, the 
optimal mapping between sensors and actuators is learned.
Learning is on the other side absent when generative principles
are implemented and studied \citep{gros2014generating}. Short-term 
synaptic plasticity, a transient form of mostly presynaptic neural 
plasticity \citep{hennig2013theoretical}, has been shown in this context 
to generate limit cycles \citep{toutounji2014behavior}
and chaotic attractors \citep{martin2016closed}. Other examples
are the homeostatic principles regulating the average neural 
activity \citep{linkerhand2013generating}, which
have been shown to induce surprisingly complex locomotive 
patters \citep{sandor2015sensorimotor}.

The control of wheeled robots, e.g.\ with skid 
steering \citep{kozlowski2004modeling}, is well 
established, with explicit mathematical models
\citep{das2006simple} being often the basis for
the regulation of either the angular velocity of the 
individual wheels \citep{jimenez2012neuro}, 
or of the respective torque 
\citep{mandow2007experimental}. One may obtain 
the sensorimotor mapping relevant for the 
neuromorphic robot at hand also by training
large neural networks \citep{conradt2015trainable}.
As an alternative, we consider here exceedingly simple
neural control schemes that are based on physical
principles and not on adaptive learning. Typically,
we need just one or two neurons per actuator.

Our starting point is the observation that
neural activity, such as the spiking rate
$y$, has a defined but limited range, say
$y\in[0,1]$. The activity of an output motor 
neuron could therefore be mapped, 
in principle, directly to the target angular 
velocity $\omega$ of the wheel, e.g.\ via 
$\omega\sim (2y-1)$. Forward and backward
motions would correspond in this setting to
distinct neural activity patterns. We examine
here in contrast a neural controller for
which the time reversal symmetry between
forward- and backward motion is broken
spontaneously under the influence of 
initial conditions.

Our robots are equipped with two active wheels 
and a third passive support wheel. A maximum of
two neurons per wheel generate self-organized locomotion,
which is both compliant and variable. A
simulated transmission rod is used to map the 
forth-and-back motion of the neural activity 
level $y\in[0,1]$ to the motor command in a 
manner that mirrors the transmission mechanism
used by traditional steam locomotives to transmit 
the force generated by the pressurized steam 
piston to the rotating wheel.

Both computer simulations and experiments with real 
robots are performed in order to assess the feasibility 
of the proposed control mechanism. We find that locomotion
is generated robustly both for individual robots and
for trains of passively coupled two-wheeled cars.
The only sensory information driving the neural activity
is propriosensoric, namely the current angle of 
the wheel the neuron controls. Cross-wheel information 
exchange is absent. Highly complex behavioral patterns 
(such as forward and backward locomotion, exploratory 
chaotic motion) 
emerge nevertheless upon interaction with the environment, 
which modifies the attractor landscapes
of the individual wheels.

Experimenting with additional top-down control signals 
we find that it is possible to kick individual wheels 
from one attractor into the basin of attraction of another 
attracting state. The self-organized limit cycles and chaotic 
attractors forming in the sensorimotor loop may hence be used
also as motor primitives.

The rest of the paper is structured as follows. 
In the Materials and Methods section the concept of kick control 
is introduced and defined mathematically, followed 
by the description of the proposed controller and 
of the experimental setup. In the Results section three 
possible implementations of kick control and their 
reliability tests are presented as a proof of concept.
The experimental findings are then investigated 
via a simple analytic model as well. Finally, a summary 
is given in the Discussion section.
\begin{figure}[t]
\centering
\includegraphics[height=0.35\textwidth]{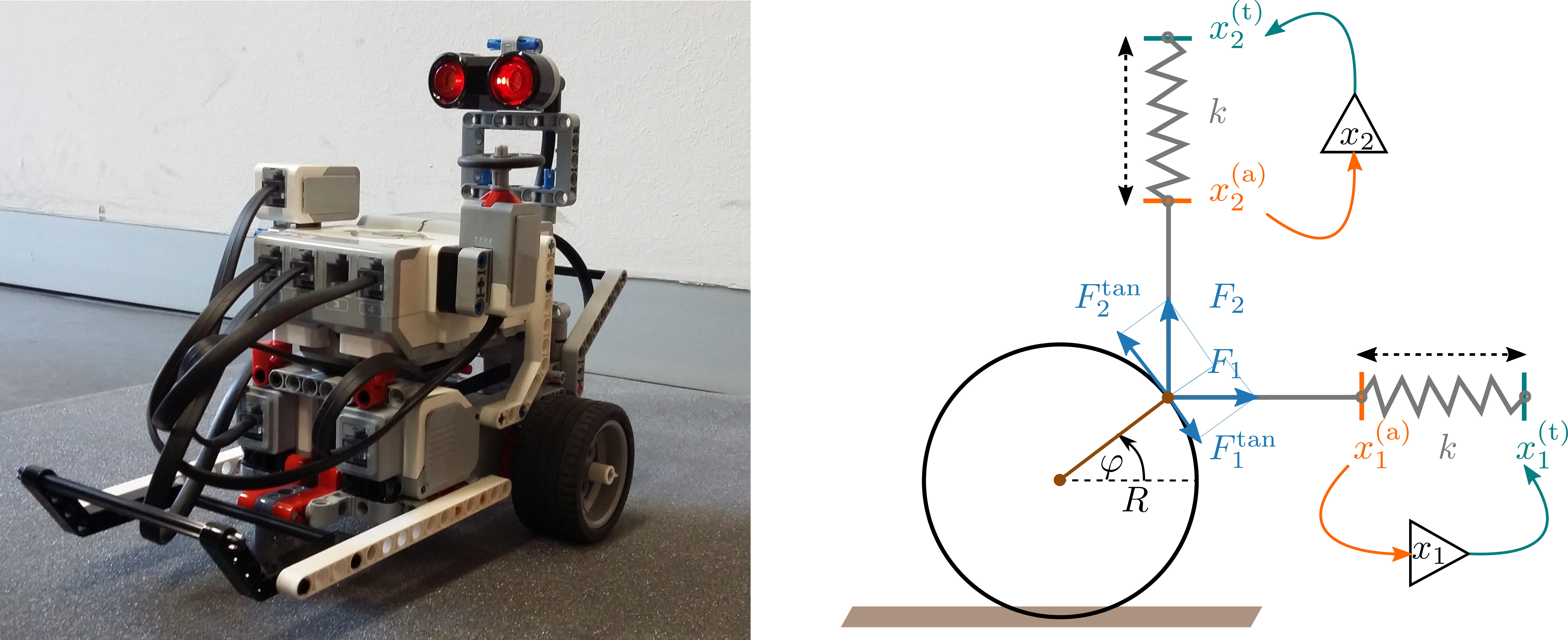}
\caption{
\textit{Left:} The two-wheeled real robot constructed with
the LEGO Mindstorms package. The active wheels are controlled 
by independent motors. A third passive wheel (the spherical 
shaped metallic wheel below the robots) keeps the body horizontal. 
\textit{Right:} A sketch of a wheel with two 
perpendicular actuators 
(\href{http://itp.uni-frankfurt.de/~gros/Movies/wheelRobot/sketch_steam_engine.mp4}
      {click for movie}).
A spring with spring constant $k$ pulls the rod (in grey) 
towards the target position $x^{(t)}_i$ (green).
The target position $x^{(t)}_i$ is determined by the output of a 
controlling neuron, as described by Eq.~(\ref{eq:target_pos}), 
which receives in turn the actual position $x^{(a)}_i$ (orange) 
of the wheel as an input. Compare Eq.~(\ref{eq:actual_pos}). 
The final torque acting on the wheel is given by the sum 
of the tangential components of the spring 
forces $F_1$ and $F_2$, see Eq.~(\ref{eq:F_tan}).
}
\label{fig:sketch_robot}
\end{figure}

\section{Materials and Methods}

From a general perspective we are interested in attracting 
states that form in the sensorimotor loop. Defining with
$\mathbf{x}_R$ and $\mathbf{x}_E$ the dynamical variables 
of the robot (R) and of the environment (E) we have
\begin{equation}
\dot{\mathbf{x}}_R \,=\,
\mathbf{f}_R(\mathbf{x}_R,\mathbf{x}_E;\mathbf{P}_R),
\qquad\quad
\dot{\mathbf{x}}_E \,=\,
\mathbf{f}_E(\mathbf{x}_R,\mathbf{x}_E;\mathbf{P}_E)
\label{dot_generic_robot}
\end{equation}
for the combined dynamics, where $\mathbf{P}_R$ and 
$\mathbf{P}_E$ parametrize respectively the time evolution of the 
robot and of the environment. Parameters distinguish 
themselves in our notation from variables in the respect 
that they change either only very slowly, as the result of 
a separation of time scales, or via actions that can be 
considered external. Control signals will modify the 
evolution equations (\ref{dot_generic_robot}), which 
describe as such locomotion generated autonomously within 
the sensorimotor loop. 

\subsection{Kick control}\label{subsect_kick_control_def}

Locomotion is characterized typically by timescales 
of seconds. One speaks of ``kick control'', when 
a robot is subjected to control sequences that are 
shorter than the time needed to complete a movement,
e.g.\ of the order of 50-200\,ms. Kick control is 
functionally dependent on the existence of multiple 
attracting states in the sensorimotor loop that 
correspond to distinct locomotive patterns. The 
control signal then serves ``to kick'' the dynamical 
system (\ref{dot_generic_robot}) into the basin of attraction 
of the desired attracting state. There are two mutually
not excluding venues.

\begin{itemize}
\item[--] ``Frozen'' kick control is present when
          the control pulse $\Delta\mathbf{x}_R$ 
          acts via
\begin{equation}
\mathbf{x}_R\ \to\ \mathbf{x}_R+\Delta\mathbf{x}_R
\label{frozen_kick_kontrol}
\end{equation}
          exclusively onto the variables $\mathbf{x}_R$ 
          of the robot. The parameters $\mathbf{P}_R$
          are not changed, they remain frozen. The
          state of the system is kicked here from its present
          state $\mathbf{x}_R$ to a new state, viz to
          $\mathbf{x}_R'=\mathbf{x}_R+\Delta\mathbf{x}_R$.
          A sudden change of parameters as in (\ref{frozen_kick_kontrol})
          corresponds to an additional strong temporary 
          force within the right-hand-side of the corresponding
          Newton equation of motion (i.e.\ to a kick).

\item[--] ``Quenched'' kick control is realized when
          the control pulse $\Delta\mathbf{P}_R(t)$, which may 
          be something like a rectangular pulse or a broadened 
          $\delta$-function, leads via
\begin{equation}
\mathbf{P}_R\ \to\ \mathbf{P}_R+\Delta\mathbf{P}_R(t)
\label{quenched_kick_kontrol}
\end{equation}
          to a sudden but transient change of the parameters 
          $\mathbf{P}_R$ of the robot. The near instantaneous 
          change of parameters catapults the system into a quenched 
          configuration for which the attracting states of 
          evolution equations (\ref{dot_generic_robot}) are morphed.
\end{itemize}

Quenched kick control will need in general a somewhat longer
control signal. The reason is that the time evolution under
the influence of the morphed attractors, that are present 
over the duration of the control pulse for the case of quenched 
kick control, needs to progress to a point in which the system
finds itself within the basin of attraction of a different
attractor once the control pulse ceases.

We have defined control signals as changes of either the 
variables or of the parameters of the robot. We may consider
alternative events that lead to changes of the state of
the environment, notably of the parameters
$\mathbf{P}_E$. This happens in particular when the robot
interacts with other objects, e.g.\ when it bounces against
a wall. The resulting transition to a different attracting
state may hence also be described at times within the terminology 
of kick control.

\subsection{Simulated steam-locomotive actuator}
\label{sec:steam}

We consider robots for which the active wheels are controlled 
independently by simple proprioceptual rate-encoding neurons. 
A finite angular velocity is attained when the internally 
generated torque interacts with the external response resulting
from friction forces, gravity and inertia. 

Neural activity covers a finite range, which can be normalized
to the interval $[0,1]$. A straightforward route for translating
the neural activity $y_i$ to a rotational mode would be to take
$y_i$ to be directly proportional, in the spirit of direct
control, to a target angular velocity. Here we consider an
alternative mechanism which allows for the generation of
self-stabilizing attractors in the sensorimotor loop. For 
this purpose we use two steam-locomotive-like actuators that
allow to translate the finite range of neural activity into
a rotational model.

The two simulated actuators used for each wheel  have
a perpendicular alignment  that allows for a continuous 
tracking. The respective transmission rods are 
fixed at one point of the perimeter of the wheel, as illustrated 
in Fig.\,\ref{fig:sketch_robot}, being moved at the other 
end by ideal springs with constant $k$. The spring forces 
\begin{equation}
F_i = k\left(x_i^{\text{(t)}}-x_i^{\text{(a)}}\right)\,,
\qquad\qquad
i=1,2\,,
\label{eq:F_i}
\end{equation}
are proportional to the distance $x_i^{\text{(t)}}-x_i^{\text{(a)}}$
between the normalized target position $x_i^{\text{(t)}}$ and  the
actual position $x_i^{\text{(a)}}$ of the wheel. The actual positions 
$x_i^{\text{(a)}}\in[-1,1]$ are determined in turn by the 
projections of the rotational angle $\varphi$ of the wheel,
measured respectively relative the horizontal and the vertical
direction:
\begin{equation}
x_1^{\text{(a)}} = \cos\varphi\,,
\qquad\qquad
x_2^{\text{(a)}} = \sin\varphi\,.
\label{eq:actual_pos}
\end{equation}
The target positions $x_i^{\text{(t)}}\in(-1,1)$ 
are provided on the other side by the output of 
two independent rate-encoding neurons,
\begin{equation}
x_i^{\text{(t)}} = 2y(x_i)-1\,,
\qquad\qquad
y(x)=\frac{1}{1+e^{-ax}}\,,
\label{eq:target_pos}
\end{equation}
which are characterized by a sigmoidal transfer function 
$y(x)\in[0,1]$ with slope $a/4$. The dynamics of the membrane 
potential $x_i$ is driven in turn by the proprioceptual input 
$x_i^{\text{(a)}}$,
\begin{equation}
\tau\dot{x}_i  = x_i^{\text{(a)}}-x_i\,,
\qquad\qquad
i=1,2\,,
\label{eq:dot_x}
\end{equation}
where the internal time scale $\tau$ is of the order 
of a few hundred milliseconds. The total tangential 
force acting on the wheel is then the vectorial sum 
of the projections of individual spring forces 
(see Fig.\,\ref{fig:sketch_robot}):
\begin{equation}
F^{\text{tan}} = F^{\text{tan}}_1 + F^{\text{tan}}_2 = 
F_1\sin\varphi - F_2\cos\varphi\,.
\label{eq:F_tan}
\end{equation}
The proposed controller translates hence the forth 
and back dynamics of the normalized neural activity 
$y\in[0,1]$ to a rotational motion for the wheel.
For an animated illustration of the steam-locomotive 
controller see the Supplementary Material. 

Note that a single actuator, e.g.\ a single horizontal 
transmission rod, would lead to a sinusoidal force 
$\propto\!F^{\text{tan}}_1$ vanishing at $\varphi=0$ and 
$\varphi=\pi$. The combined force (\ref{eq:F_tan})
is on the other hand always finite when two actuators 
with a perpendicular alignment are employed. See
Fig.\,\ref{fig:sketch_robot} and the illustrating movie
presented in the supplementary material.

\subsection{LEGO Mindstorms robot}
\label{subsect_LEGO_Mindstorms_robot}
 
To test the presented actuators we constructed robots with 
two active wheels using the LEGO Mindstorms Education
Core Set (see the left picture of Fig.\,\ref{fig:sketch_robot}). 
For more details about the experimental setup see the Supplementary 
Material. A third passive wheel keeps the body of the 
robot in a horizontal position. The active wheels are driven
by motors that provide sensory feedback regarding the
angle $\varphi$ of the individual wheels. The working regime
of the LEGO motors is finite, that is they respond to
inputs $\tilde{M}\in[-\tilde{M}^\mathrm{max},\tilde{M}^\mathrm{max}]$.
In order to comply with this constraint we mapped
the simulated tangential force $F^\mathrm{tan}$, defined by 
Eq.~(\ref{eq:F_tan}), via
\begin{equation}
\tilde{M} = \tilde{M}^\mathrm{max}\tanh \left(F^\mathrm{tan}\right)
\label{eq:torque}
\end{equation}
to the motor signal $\tilde{M}$. For an elastic response 
we used typical relative motor commands of the order of
$M=\tilde{M}/\tilde{M}^\mathrm{max}\in[-0.7,0.7]$.
Absolute time was measured at the start of every control 
loop and compared with the last time the
control loop was called. The such determined
time difference $\Delta t$ between two successive
instances of the control loop was used for solving 
(\ref{eq:dot_x}) via a straightforward Euler integration. 
On the average we had $\Delta t\approx 40\,\mathrm{ms}$.

The motor stalls for low motor powers, viz when
$|\tilde{M}|\le 0.1\tilde{M}^\mathrm{max}$, as a 
consequence of the internal friction of the gearing.
A minimal torque is hence required for the robot to
start moving. The overarching dynamical system, 
describing the body, the internal controller and the
interactions between body and environment, allows 
for the generation of self-organized attractors
\citep{sandor2015sensorimotor} that correspond
to different motion patterns, the motor primitives
\citep{ijspeert2002learning}. 

\begin{figure}[t]
\centering
\includegraphics[width=1.0\textwidth]{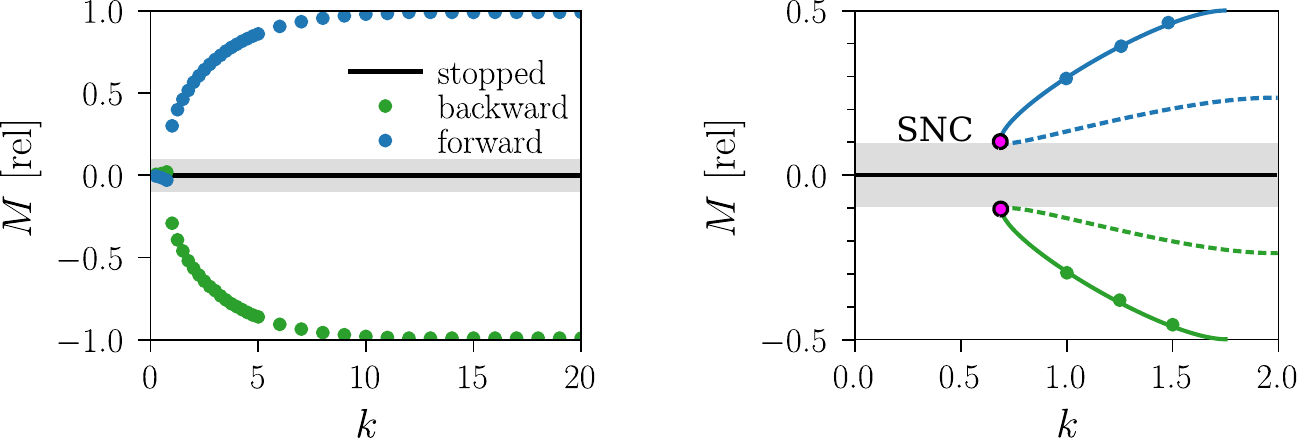}
\caption{Motion primitives corresponding to non-moving
and to forward and backward limit-cycle locomotion. Note
that the motor stalls for $|M|<M^\mathrm{thr}\approx0.1$
(grey area), viz when the torque $M$ is unable
to overcome the internal friction.
\textit{Left}: The measured output torque $M$  
for $a=4$ and $\tau=250\,\mathrm{ms}$ as a 
function of the spring constant $k$. The 
measurements are performed after the robot
settles in the forward (blue) or in the 
backward (green) attractor. For small $k$ the 
robot stops moving and the tangential force 
vanishes.
\textit{Right}: An enlargement of the region
$k\in[0,2]$ showing the measurements (blue/green dots)
and the corresponding stable limit cycle (blue/green
lines). A saddle-node bifurcation of limit-cycles 
(SNC) is likely to occur when the torque is counteracted 
by the internal friction. The resulting unstable limit 
cycles (dashed blue/green lines) are shown. 
}
\label{fig:locomotion_attractors}
\end{figure}

\section{Results}

\subsection{Self-organized attractors as motor primitives}

In the normal mode the robot disposes, as shown in 
Fig.\,\ref{fig:locomotion_attractors}, of three possible states:
stopped, forward and backward moving. For the
forward and backward limit cycle locomotion the
torque acting on the wheels is quasi-stationary,
an observation that is consistent with the analytic
treatment detailed out in Sect.\,\ref{sec_analytic_modelling}.
Additionally, the robot may also rotate around its own axis, 
which happens when the two active wheels turn in opposite
directions or when only one of the two wheels turns.

The measured speed of the robot in the forward and 
backward moving modes is $v=0.35\,\mathrm{m/s}$ 
for $k=8$, $a=4$ and $\tau=250\,\mathrm{ms}$, a
setting that makes use of about 80-90\% of the maximal 
power of the motor. In addition to the basic limit cycle 
attractors one finds for $a=4$, $k=15$ and 
$\tau=1000\,\mathrm{ms}$ a chaotic attractor, for which 
the motors switch irregularly between the destabilized 
fundamental modes of the individual wheels, as
illustrated in Fig.\,\ref{fig:chaotic_attractors}. For 
a video of the chaotic dynamics of the robot see the 
Supplementary Material. We did not attempt, for the
chase of the LEGO robot, to fully map the set of parameters 
for which a stable chaotic attractor exists, e.g.\
by evaluating the Lyapunov exponents in the 
context of navigation \citep{harter2005chaotic}.
We note that chaotic attractors have been shown 
to be useful in the design and construction of 
spatial navigation models \citep{voicu2004spatial}.

\begin{figure}[t]
\centering
\includegraphics[width=1.0\textwidth]{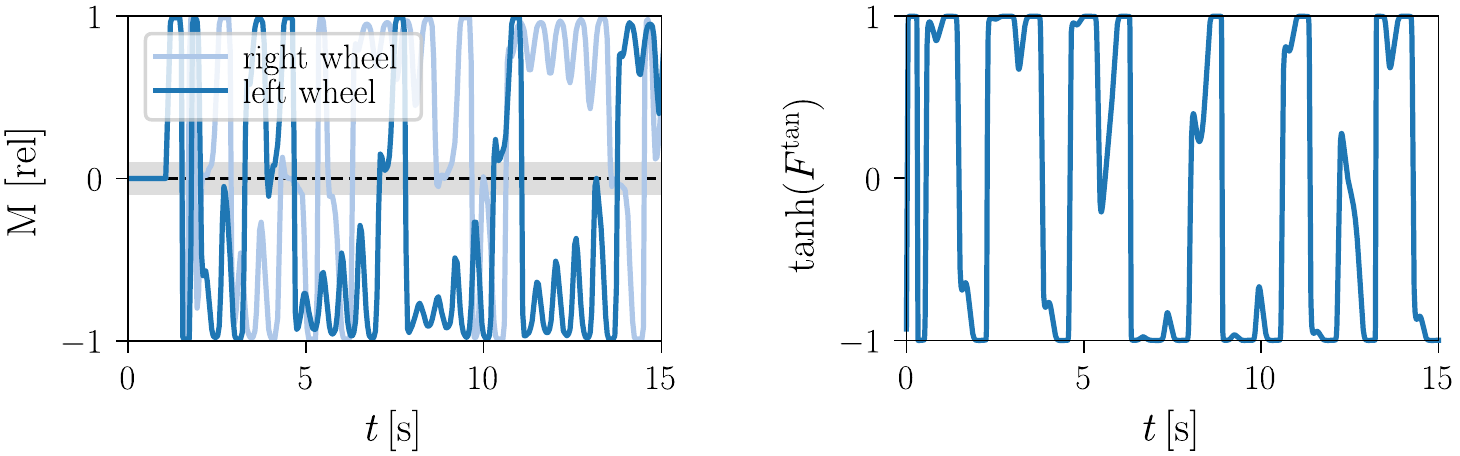}
\caption{
Time series of the motor torques (\ref{eq:torque})
in the chaotic mode, in relative units,
for $a=4$, $\tau=1000\,\mathrm{ms}$ and $k=15$.
\textit{Left}: For the LEGO robot
(\href{http://itp.uni-frankfurt.de/~gros/Movies/wheelRobot/lego_chaos.mp4}
      {click for movie}). 
The light gray shaded area indicates the 
$|M|<M^\mathrm{thr}\approx0.1$ region where 
the motor stalls.
\textit{Right}: For the analytic model (\ref{eq:dot_x_phi}),
with $F^\mathrm{tan}$ given by the torque on right-hand side of
$I\dot\omega$. Here we took $f=0.5$ for the friction 
and $I=0.05$ for the moment of inertia.
}
\label{fig:chaotic_attractors}
\end{figure}
 
\subsection{Kick control for embodied robots}
 
The presence of coexisting attractors, viz of
multistability \citep{pisarchik2014control}, allows 
to switch between the basic modes without the need 
to modify the internal parameters of the system 
for the entire locomotion. The transitions between 
the individual attractors may be induced by external 
physical stimuli, such as collisions with other robots 
or with the environment \citep{martin2016closed}.

\begin{figure}[t]
\centering
\href{http://itp.uni-frankfurt.de/~gros/Movies/wheelRobot/lego_wallbounce.mp4}
{\includegraphics[height=0.3\textwidth]{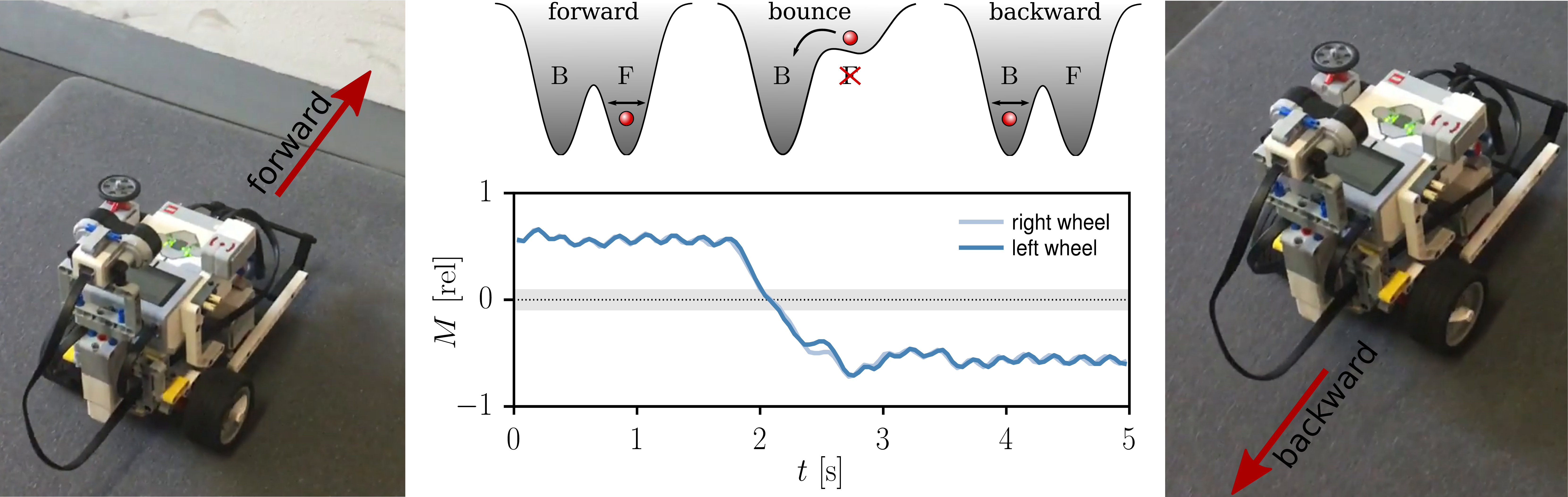}}
\caption{Collision induced switch of attractors.
\textit{Middle}: The time series of the relative
torque $M$ acting on the wheels for  
$\tau=250\,\mathrm{ms}$, $k=2$ and $a=4$.
The gray shaded region indicates the minimal 
torque needed to start the motor, $M^\mathrm{thr}=0.1$.
The superimposed sketches illustrate
a double-well potential with the minima 
corresponding to two coexisting attractors.
The phase point (red ball) stays around the minima
even in the presence of noise or small oscillations.
The system is located in the backward attractor (B)
when the robot collides with a wall and the forward 
attractor (F) is destabilized. The 
total torque $M$ changes consequently its sign.
\textit{Right and left}: The Lego robot
before and after colliding with the wall
(\href{http://itp.uni-frankfurt.de/~gros/Movies/wheelRobot/lego_wallbounce.mov}
      {click for movie}).
}
\label{fig:wall_bounce}
\end{figure}

A robot initialized in the forward moving mode 
is able to reverse direction when bouncing off 
a wall placed perpendicularly to the direction 
of locomotion, as illustrated in Fig.\,\ref{fig:wall_bounce}.
The reversal of direction is performed in this case
autonomously, that is in absence of any additional
control signals. It occurs because the forward mode 
gets destabilized for the duration of the collision,
whereas the basin of attraction of the backward 
mode expands correspondingly. The flow in phase space 
is then drawn towards the backward attractor, where it 
stays after the forward attractor reemerges upon pulling 
away from the wall. Autonomous switching between forward 
and backward modes when colliding with obstacles occurs
robustly, as we demonstrated by a series of experiments 
on rough surfaces (see Supplementary Video 3). 

An alternative possibility to generate switches 
between coexisting attractors is to kick the phase 
point of the dynamical system to the basin of attraction 
of another attractor, termed here as kick control. This 
may be realized by applying short duration input stimuli. 
We present here three intuitive mechanisms, which may
be classified, as discussed in
Sect.\,\ref{subsect_kick_control_def}, as
`frozen' and `quenched' kick control.

\begin{figure}[t]
\centering
\includegraphics[width=0.75\textwidth]{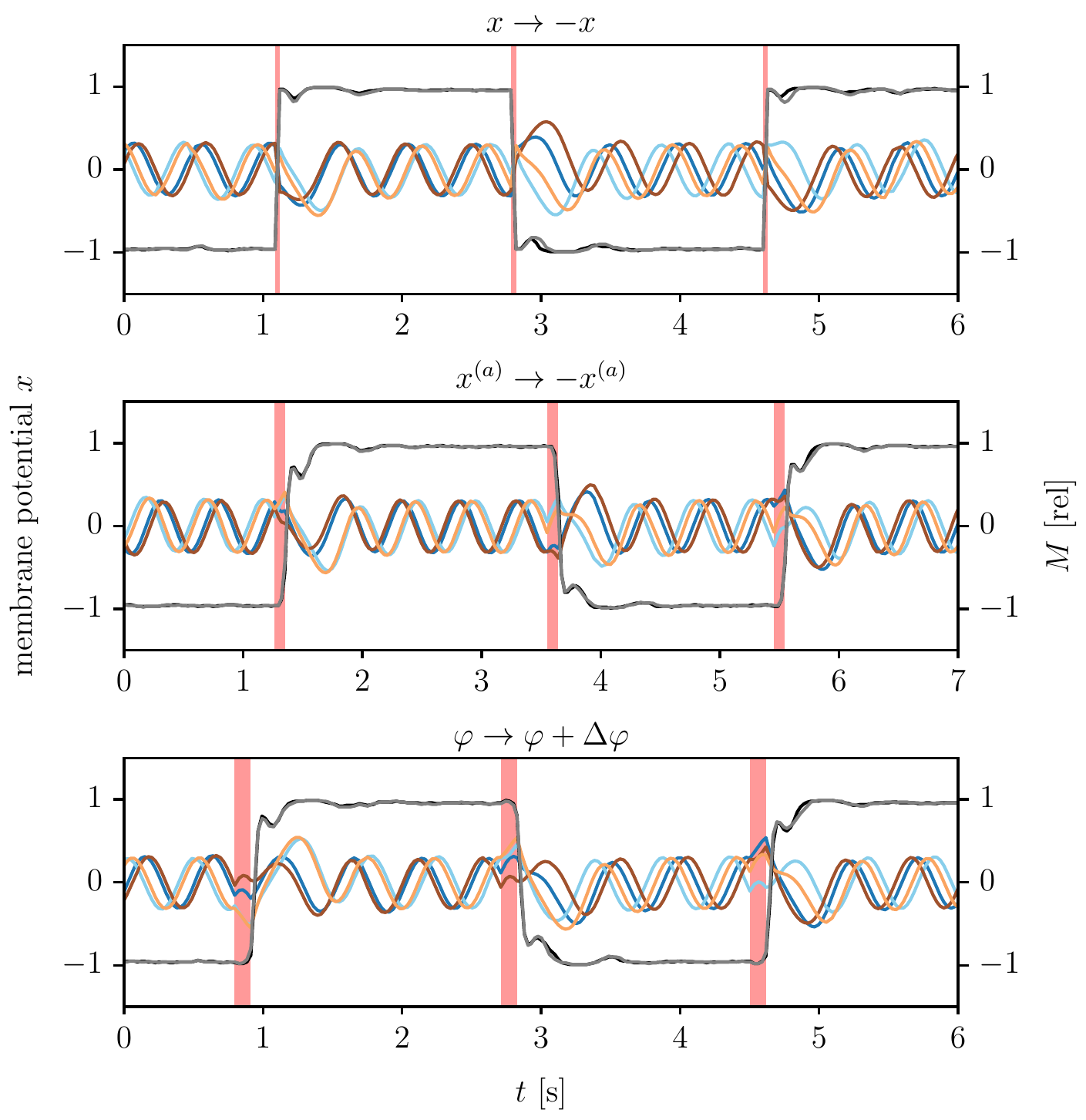}
\caption{Kick controlling the LEGO Mindstorms robot. Shown are the
time series of the membrane potentials (brown/orange and blue/cyan
lines for $x_1$/$x_2$ of the left/right wheel), together with the
normalized motor control $M=\tilde{M}/\tilde{M}^{max}$ (black/grey lines
for left/right wheel). The parameters $\tau=250\,\mbox{ms}$, $k=8$ 
and $a=4$ lead to an angular frequency of 
$\omega/(2\pi)\approx 2\,\mbox{Hz}$.
\emph{Top}: Inverting all membrane potentials $x_i$,
compare (\ref{eq:kickControl_membrane_inversion}), with
the times indicated by the vertical red lines
(\href{http://itp.uni-frankfurt.de/~gros/Movies/wheelRobot/lego_control_membrane_switch.mp4}
      {click for movie}).
\emph{Middle}: Inverting all actual positions
$x_i^{(a)}$, see (\ref{kickControl_xa_inversion}),
for three ticks of the updating cycle (roughly 90\,ms,
as indicated by the red vertical bars).
\emph{Bottom}: Adding a phase shift of 
$\Delta\varphi=\pm3\pi/4$ to the measured angle
of the wheels, see~(\ref{kickControl_delta_phi}).
The length of the control signal is here 4 ticks,
corresponding to 120\,ms. Note that the time
to settle in the reverse limit cycle exceeds
the duration of the kick signal. The relative
phase of the left and the right wheel increases
considerably after the second reversal.
}
\label{fig:kick_control_timeline}
\end{figure}

\subsubsection{Frozen kick control}

A reliable direction reversal may be induced
by inverting membrane potentials via
\begin{equation}
x_1\ \rightarrow\ -x_1,
\qquad\quad
x_2\ \rightarrow\ -x_2\,,
\label{eq:kickControl_membrane_inversion}
\end{equation}
at given time. Eq.~(\ref{eq:kickControl_membrane_inversion})
induces an instantaneous change of the internal variables
that does not affect the parameters of the one-neuron
controller. It corresponds therefore, as defined in 
Sect.~\ref{subsect_kick_control_def}, to frozen kick 
control. The respective time-series of the membrane 
potentials are shown in Fig.\,\ref{fig:kick_control_timeline}.
The reversal of the direction occurs fast, depending
however on the state the actuator was in when
the membrane potentials were inverted. It is 
furthermore noticeable, in particular following the
second kick signal, that it may take half a second
or more to fully settle into the reversed attractor.
A relative phase slip may be induced in addition
in between the two wheels, which are mechanically
not precisely identical. The here considered variant 
(\ref{frozen_kick_kontrol}) of frozen kick control 
is furthermore 100\% reliable in direction reversal 
tests (for a demonstration see Supplementary Video 4).
This is due to the fact that the sign flip of 
internal variables $x_{1,2}$ leads instantaneously 
to a motor torque of opposite direction, 
compare Eqs.~(\ref{eq:F_i}) and (\ref{eq:F_tan}).

\subsubsection{Quenched kick control}

We considered two variants of quenched kick control.
For the first variant one substitutes the actual wheel 
positions $x_{1,2}^{(a)}$ by
\begin{equation}
\begin{array}{rcl}
x_1^{(a)}\ & \rightarrow & (1-2\beta)\cos(\varphi)\\
x_2^{(a)}\ & \rightarrow & (1-2\beta)\sin(\varphi)
\end{array}\,,
\qquad\quad
\beta=\left\{\begin{array}{rl}
0 & \mbox{(control\ off)}\\
1 & \mbox{(control\ on)}
\end{array}\right.~.
\label{kickControl_xa_inversion}
\end{equation}
The new parameter $\beta=\beta(t)$ is turned on 
for a finite control period $\Delta t$, as 
illustrated in Fig.\,\ref{fig:kick_control_timeline}.
This control procedure mimics (\ref{eq:kickControl_membrane_inversion})
in the sense that the reversal of the membrane potentials
is not achieved by a direct kick in the phase space 
of internal variables, but by a change of parameters. 
Compare Eq.~(\ref{eq:dot_x}).

\begin{figure}[t!]
\centering
\includegraphics[width=0.8\textwidth]{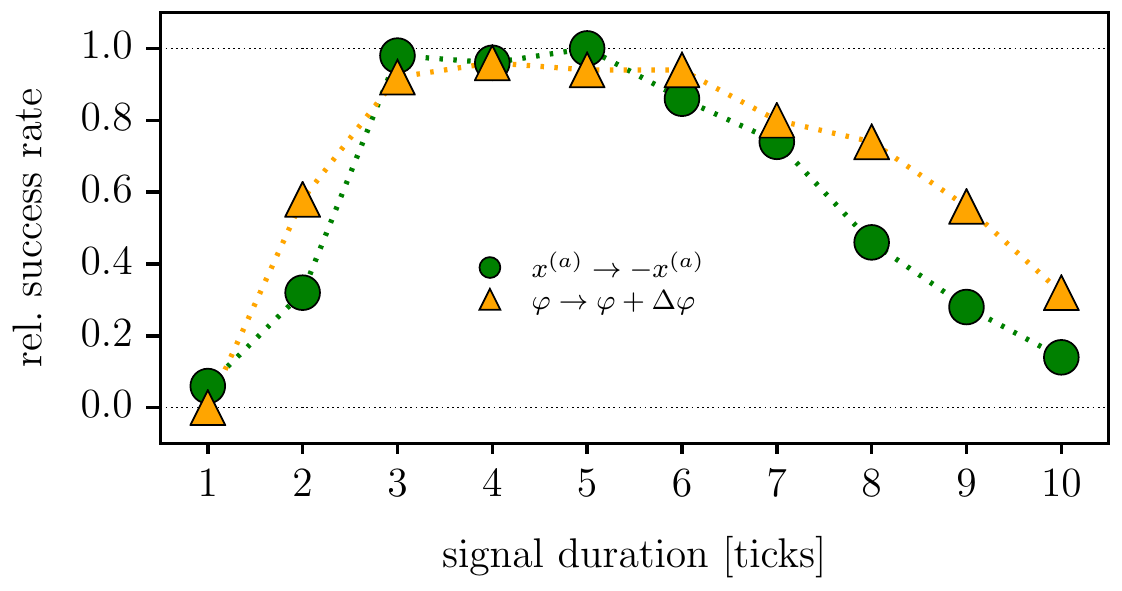}
\caption{Reliability of quenched kick control procedures 
for the LEGO Mindstorms robot in terms of the relative 
success rate when testing direction reversal. The parameters are 
$\tau=250\,\mbox{ms}$, $k=8$ and $a=4$, with each data 
point corresponding to the average of 50 trials. The 
duration $\Delta t$ of the control signal is given in 
terms of ticks, with a tick corresponding to the period 
of the control cycle of the  Lego robot (about 30\,ms).
Maximal reliability is achieved for 3-5 ticks (90-120\,ms),
both for reversing the actual position $x^{(a)}$ (compare
(\ref{kickControl_xa_inversion}), green circles) and when
adding a phase-shift of $\Delta\varphi=\pm3\pi/4$ to the
angle of the wheel (as defined by (\ref{kickControl_delta_phi}), 
orange triangles). A positive $\Delta\varphi$ induces here a
transition from forward to backward locomotion (reversely
for negative $\Delta\varphi$).
}
\label{fig:resultsKickControl}
\end{figure}

Real-world robots come with a control cycle that
discretizes time. We consequently measure the time 
$\Delta t$ during which $\beta(t)$ is active in terms
of control cycles (ticks). In Fig.\,\ref{fig:resultsKickControl}
we present the results of an experiment testing
the reliability of (\ref{kickControl_xa_inversion}),
that is for the probability that the robot reverses 
direction for a given $\Delta t$. For the time-series
shown in Fig.\,\ref{fig:kick_control_timeline} the
robot received a kick signal for three ticks,
viz for about 90\,ms.

For an alternative type of quenched kick control
we consider with
\begin{equation}
\begin{array}{rcl}
x_1^{(a)}\ & \rightarrow & \cos(\varphi+\gamma\Delta\varphi)\\
x_2^{(a)}\ & \rightarrow & \sin(\varphi+\gamma\Delta\varphi)
\end{array}\,,
\qquad\quad
\gamma=\left\{\begin{array}{rl}
0 & \mbox{(control\ off)}\\
1 & \mbox{(control\ on)}
\end{array}\right.
\label{kickControl_delta_phi}
\end{equation}
a shift in the sensory value of the angle of the
robot. The corresponding reliability statistics 
are also shown in Fig.\,\ref{fig:resultsKickControl}.
Both types of quenched kick control, as defined by
Eqs.~(\ref{kickControl_xa_inversion}) and
(\ref{kickControl_delta_phi}), need to be applied
for 3-5 control ticks, corresponding to 90-150\,ms,
for the robot to turn direction. For the experiment
presented in Fig.\,\ref{fig:kick_control_timeline}
we used 4 ticks when using kicking the robot
via (\ref{kickControl_delta_phi}).
 
\subsection{Self-organized train of cars}

\begin{figure}[t!]
\centering
\includegraphics[height=0.2\textwidth]{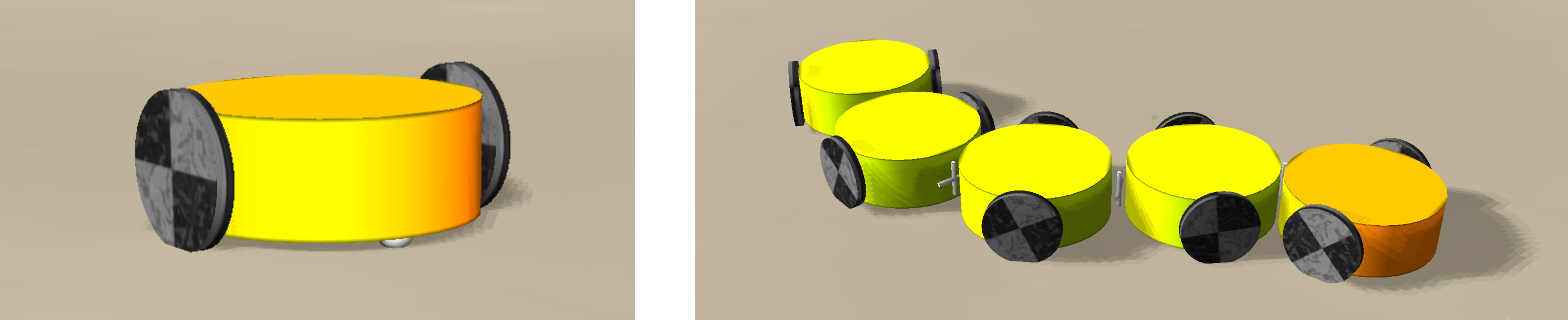}
\caption{Snapshots from the LPZRobots 
simulations environment \citep{der2012playful} 
of two-wheeled car-like and train-like robots.
\textit{Left}: The two active wheels 
(black) are controlled as described in
Sect.~\ref{sec:steam}. The passive support 
wheel (white) below the body (yellow) prevents
the car from tipping over
(\href{http://itp.uni-frankfurt.de/~gros/Movies/wheelRobot/lpz_wheeled_slopes.mp4}
      {click for movie}).
\textit{Right}: A train robot created by connecting 
cars via passive torsion springs
(\href{http://itp.uni-frankfurt.de/~gros/Movies/wheelRobot/lpz_wheeled_train.mp4}
      {click for movie}).  
}
\label{fig:LPZ_robots}
\end{figure}

We used the LPZRobots simulations environment \citep{der2012playful}
to simulate robots with two actuated wheels, as
illustrated in Fig.\,\ref{fig:LPZ_robots}. The individual
robots have a body mass of $0.1\,\mbox{kg}$, 
wheel mass of $0.05\,\mbox{kg}$, wheel radius 
of $3\,\mbox{cm}$, body radius of $10\,\mbox{cm}$ and
body height of $5\,\mbox{cm}$. The dimensionless friction 
coefficient is 0.1. The cars are controlled as 
described in Sect.~\ref{sec:steam}, but this time 
only a single simulated transmission rod is employed.
This is possible, as the motor of the simulated
robots transits to an idle state in the absence 
of an input signal.

The single two-wheel car follows intricate non-holonomic 
trajectories when put in an environment containing confining 
slopes. We also constructed trains composed of two-wheel cars 
coupled passively via torsion springs. All ten wheels,
for the case of the five-car train shown in
Fig.\,\ref{fig:LPZ_robots}, are independent. One
observes that the ten wheels coordinate
their rotations speed and direction, reacting in
a coordinated manner upon encountering objects
in structured environment. The resulting locomotion 
of the train is a prime example of a self-organizing
process (see Supplementary Videos 5 and 6).

\section{Analytic modeling}\label{sec_analytic_modelling}

The fundamental attractor of the individual wheels
may be studied analytically when lumping the
feedback of the environment into a single
equation of motion that contains friction in
terms of a friction force $\propto\!f\omega$. The 
resulting dynamical system is then
\begin{equation}
\begin{aligned}
\tau\dot{x}_1  &= \cos\varphi-x_1\\
\tau\dot{x}_2  &= \sin\varphi-x_2\\
\dot{\varphi}  &= \omega\\
I\dot{\omega}   &= k\big(2y(x_1)-1\big)\sin\varphi
                       -k\big(2y(x_2)-1\big)\cos\varphi - f\omega\,,
\end{aligned}
\label{eq:dot_x_phi}
\end{equation}
where $\omega$ denotes the angular velocity and $\varphi$
the angle of the wheel. The membrane potential of
the horizontal and vertical controllers are,
as illustrated in Fig.\,\ref{fig:sketch_robot},
$x_1$ and $x_2$. The moment of inertia of the wheel 
is proportional to $I$, the spring constant by $k$,
the friction coefficient by $f$, the time constant 
of the membrane potentials by $\tau$ and the 
transfer function $y(x)$ of the controlling neurons 
by the sigmoidal $y(x)=1/(1+\exp(-ax))$.

Note that $x_1$ and $x_2$ are internal variables of the
robot, whereas $\varphi$ (and consequently also $\omega$)
corresponds to the physical angle of the wheel and
therefore to an environmental variable. This is because the
body of the robot, including the wheels, are, from the
perspective of the neural circuitry, part of the environment.
We also point out that the motor power needs to exceed a
certain threshold for the LEGO Mindstorms actuators 
to become active, as explained in 
Sect.~\ref{subsect_LEGO_Mindstorms_robot}. This feature 
goes however beyond Eqs.~(\ref{eq:dot_x_phi}).

\subsection{Stationary wheel}

The system of four coupled differential 
equations~(\ref{eq:dot_x_phi}), possesses eight
trivial fixpoints, characterized by
\begin{equation}
\omega^*_n=0, \qquad\qquad
\varphi^*_n=n\pi/4, \qquad\qquad
n\in\{0,\dots,7\}~,
\label{eq:trivial_fixpoints}
\end{equation}
of which the odd multiples of $\pi/4$ are always 
unstable. The even multiplies of $\pi/4$ are on 
the other side stable/unstable for small and large ratios 
of $ak\tau/(2f)$, respectively, as we will show further
below. The two-wheeled robot can hence be in 
$4\times4=16$ non-moving states corresponding to the 
16 combinations of stable fixpoints of 
Eq.~(\ref{eq:trivial_fixpoints}) of the left and right wheel. 

\subsection{Homoclinic route to locomotion}
\label{Sect_Route_locomotion}

\begin{figure}[t!]
\centering
\includegraphics[width=1.0\textwidth]{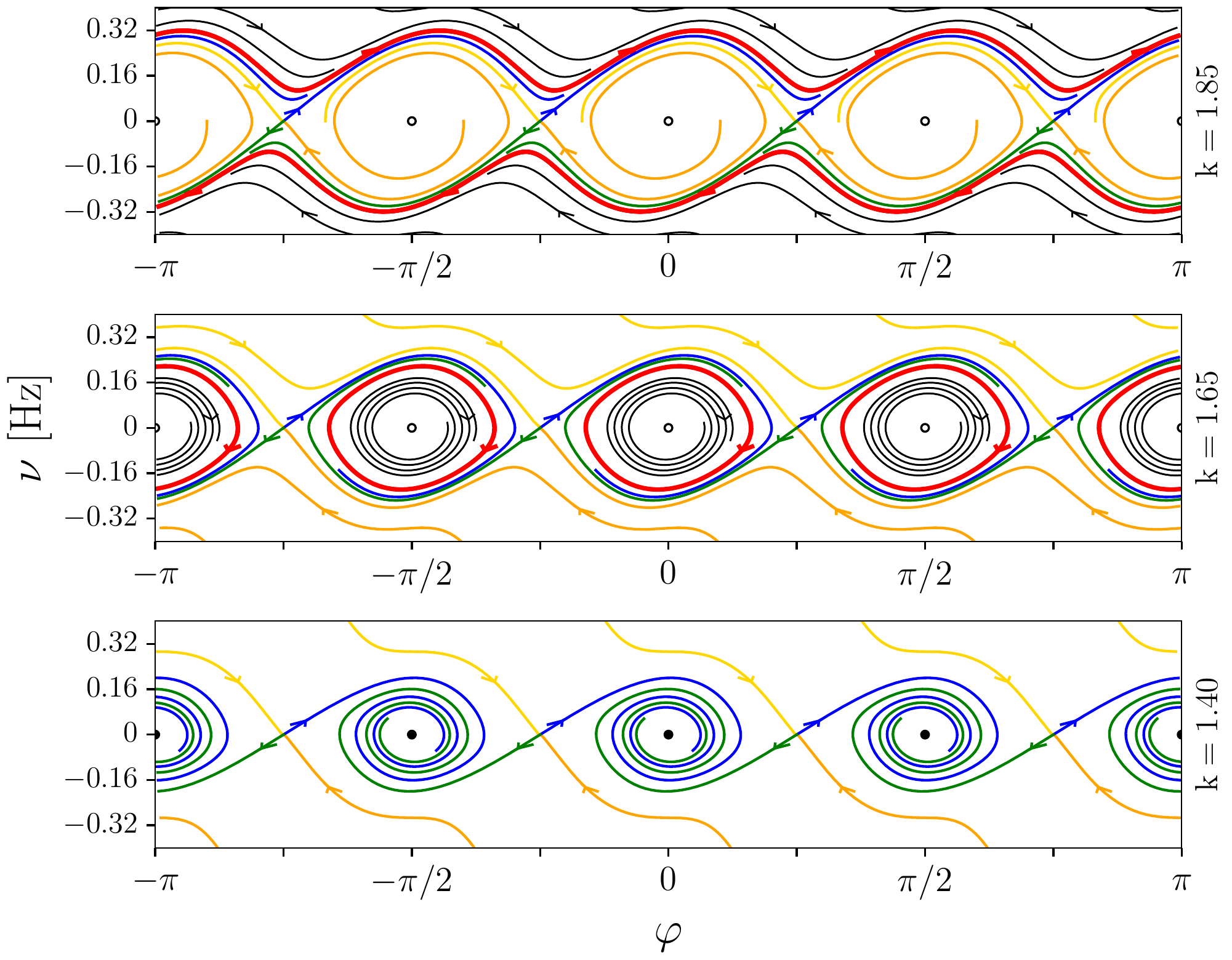}
\caption{Illustration of the homoclinic route to 
locomotion discussed in Sect.~\ref{Sect_Route_locomotion}.
The phase-space plots $(\varphi,\nu=\omega/(2\pi))$ 
are for the analytic model (\ref{eq:dot_x_phi}), with
$a=4$, $\tau=0.25$, $I=0.05$ and $f=0.5$. The spring 
constant is $k=1.85/1.65/1.4$ (top/middle/bottom).
The fixpoints located at odd multiples of $\pi/4$
are saddles. Shown are stable (yellow/orange) and 
unstable (blue/green) manifolds, limit cycles (red),
stable and unstable foci (filled/open circles)
and some selected generic trajectories (black).
\textit{Top:} Limit-cycle locomotion, with the
stable orbit winding around $\varphi\in[-\pi,\pi]$.
\textit{Middle:} Forth-and-back rolling with
the angle $\varphi$ of the wheel being limited to
a finite range around multiples of $\pi/2$.
\textit{Bottom:} Stationary states with 
$\varphi\to n\pi/2$. 
}
\label{fig:homoclinicTransition}
\end{figure}

In order to understand the transition from
the fixpoint solutions (\ref{eq:trivial_fixpoints})
to locomotion we use
\begin{equation}
\cos\varphi(t')\approx \cos\varphi(t)-\sin\varphi(t)\dot{\varphi}(t)(t'-t)
= \cos\varphi(t)+\sin\varphi(t)\omega(t)(t-t')~,
\label{phi_t_prime_approximation}
\end{equation}
which is valid for $\omega\tau\ll1$ and $t-t'\ll\tau$, 
to expand the formal integral 
\begin{equation}
x_1(t)=\frac{1}{\tau}\int_{-\infty}^t dt'\cos\varphi(t')\,e^{-(t-t')/\tau}
\approx \cos\varphi(t)+\omega(t)\tau\sin\varphi(t)
\qquad\quad
\label{x_1_integral}
\end{equation}
of $x_1(t)$. An
equivalent expansion may be derived for $x_2(t)$.
One next expands the neural transfer functions
occurring on the right-hand side of $\dot\omega$,
\begin{equation}
\begin{array}{rcl}
y(x_1) &\approx& y(\cos\varphi)+ay(\cos\varphi)(1-y(\cos\varphi))\omega\tau\sin\varphi
\\
y(x_2) &\approx& y(\sin\varphi)-ay(\sin\varphi)(1-y(\sin\varphi))\omega\tau\cos\varphi
\end{array}
\label{y_omega_tau_expansion}
\end{equation}
for small $\omega\tau$. The equations of motion
(\ref{eq:dot_x_phi}) then take the form 
\begin{equation}
\dot\varphi = \omega,\qquad\quad
I\dot\omega = F(\varphi)+\gamma(\varphi)\omega~,
\label{dot_phi_omega}
\end{equation}
where $F(\varphi)$ is a mechanical force and
$\gamma(\varphi)$ the coefficient of an 
adaptive friction. The respective expressions are
\begin{equation}
F= 2k\big[ y(\cos\varphi)\sin\varphi
          -y(\sin\varphi)\cos\varphi\big]
  +k\big[\cos\varphi-\sin\varphi\big]
\label{force_F}
\end{equation}
and
\begin{equation}
\gamma = 2ka\tau\big[
y(\cos\varphi)(1-y(\cos\varphi))\sin^2\varphi
+y(\sin\varphi)(1-y(\sin\varphi))\cos^2\varphi
\big]-f~.
\label{friction_gamma}
\end{equation}
Within this approximation, 
one finds that $\gamma(\varphi)$ is negative for 
all $\varphi\in[0,2\pi]$ when
\begin{equation}
k<k_c,\qquad\quad k_c=\frac{2f}{a\tau}~.
\label{k_critical}
\end{equation}
The system is  purely dissipative when $\gamma<0$, which 
implies that the fixpoints $\varphi_{2n}^*=n\pi/2$ are 
stable for $k<k_c$. Locomotion is next achieved via a two-step 
process, as illustrated in Fig.~\ref{fig:homoclinicTransition},
when increasing the spring constant $k$ beyond $k_c$.
\begin{itemize}
\item[--] The fixpoints $\varphi_{2n}^*=n\pi/2$ undergo 
          a supercritical Hopf bifurcation at $ak\tau=2f$, 
          viz when $\gamma(\varphi_{2n}^*)$ becomes positive.
          The angle $\varphi$ of the wheel then oscillates
          around the previously stable fixpoint 
          $\varphi_{2n}^*$, with a trajectory 
          that corresponds to a periodic forth-and-back
          motion of the robot.

\item[--] The amplitude of the limit cycle in $\varphi$
          will reach eventually, when $k$ is further 
          increased, the saddles at $\varphi_{2n}^*=n\pi/2+\pi/4$.
          The limit cycle will then merge with the respective
          stable and unstable manifold of the saddle and undergo
          a Taken-Bogdanov-type \citep{gros2015complex} 
          homoclinic bifurcation. Above this transition 
          the four symmetry-related limit cycles around 
          $\varphi_{2n}^*=n\pi/2$ merge into a large cycle.
          The wheel then performs complete rotations.
\end{itemize}

We note that an equivalent merging of symmetry related
limit cycles across a global bifurcation has been observed
in a study of prototype dynamical systems \citep{sandor2015versatile}. 

\subsection{Constant velocity approximation}

The angular moment $\omega$ becomes nearly constant
for $k$ far above the infinite-period transition.
The solution of Eq.~(\ref{eq:dot_x}) obtained in
the limit $t\rightarrow\infty$ is given for the case of
a constant angular velocity as 
\begin{equation}
x_1(t)=\frac{\cos(\omega t) + \omega\tau\sin(\omega t)}{1 + \omega^2\tau^2},
\qquad\quad
x_2(t)=\frac{\sin(\omega t) - \omega\tau\cos(\omega t)}{1 + \omega^2\tau^2}\,,
\label{eq:x-t}
\end{equation}
where we have used $\varphi(t)=\omega t$.
Assuming small amplitude oscillations for the 
membrane potential, $ax_i\ll1$, we can linearize
the transfer function $y(x)$ around $y(0)=1/2$.
The total tangential force $F^{\tan}$ defined by 
Eq.~(\ref{eq:F_tan}) then becomes constant,
\begin{equation}
F^{\tan} = 
\frac{a k \omega \tau/2}{1+\omega^2\tau^2},
\qquad\quad
\frac{a k \omega \tau/2}{1+\omega^2\tau^2}=f\omega,
\qquad\quad
\omega^*_{\pm}=\pm\sqrt{\frac{a k\tau -2f}{2f\tau^2}}~,
\label{eq:F_tan_ana}
\end{equation}
where the second equation corresponds to the balance
between $F^{\tan}$ and the friction force $f\omega$ in 
Eq.~(\ref{eq:dot_x_phi}). 
Locomotion vanishes in the constant-$\omega$
approximation for $k<k_c=2f/(a\tau)$, viz at the 
critical spring constant $k_c$ defined in (\ref{k_critical}).

The two symmetrical branches corresponding to the 
stable attractors of forward and backward motion 
can be seen in the experimentally constructed 
bifurcation diagram shown in the right panel 
of Fig.\,\ref{fig:locomotion_attractors}. The internal friction 
forces of the motor induces in addition two symmetry-related
saddle-node bifurcations of limit cycles, when reducing
$k$, such that the torque $M$ drops discontinuously 
to zero as $k\rightarrow k_\mathrm{c}$. The internal
threshold of real-world motors impacts the route
to chaos hence qualitatively. Compare 
Sect.~\ref{Sect_Route_locomotion}.

\subsection{Numerical and analytic phase diagram}

\begin{figure}[t!]
\centering
\includegraphics[width=1.0\textwidth]{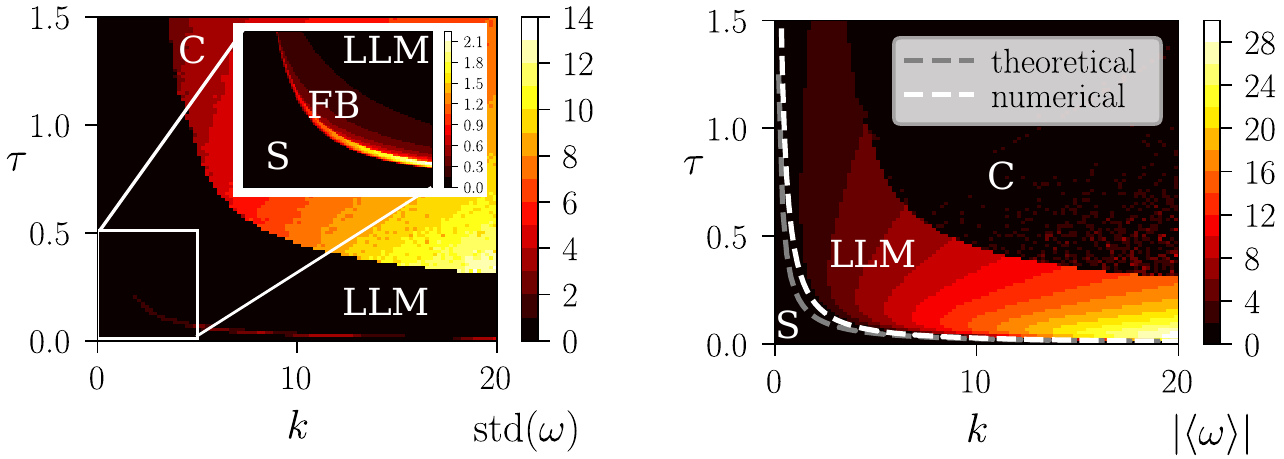}
\caption{
\textit{Left:} The numerical phase diagram, as obtained by 
simulating (\ref{eq:dot_x_phi}) for $a=4$, $I=0.05$ and 
$f=0.5$. Shown is the standard deviation
$\mathrm{std}(\omega)=\sqrt{\langle\omega^2\rangle-\langle\omega\rangle^2}$
of the angular frequency $\omega$ of the wheel (color coded),
as averaged over time. Low values of $\mathrm{std}(\omega)$ correspond
to vanishing or near constant $\omega(t)$, as for the fixpoint
solution (S) and for the limit-cycle locomotion (LLM). The
transition to chaos (C) takes place through a crisis,
with the route to locomotion occurring, as discussed in
Sect.~\ref{Sect_Route_locomotion}, via an intermediate
phase of forth-and-back rolling (FB). The inset 
enlarges the region $k\in[0,5]$ and $\tau\in[0,0.5]$. 
\textit{Right:} The average angular frequency 
$|\langle\omega\rangle|$ (color coded).
The dashed lines correspond to the instability lines of the
fixpoint solution, as obtained numerically (white) and
analytically (grey), where the theoretical result is 
$k_c\tau_c=2f/a$. See Eq.~(\ref{k_critical}).
}
\label{fig:phaseDiagram}
\end{figure}

In Fig.~\ref{fig:phaseDiagram} we present the
phase diagram, as obtained by integrating
(\ref{eq:dot_x_phi}) numerically. Four
phases are found, a stationary fixpoint
phase (S) at low values of $k\tau$, a phase
corresponding to forth-and-back (FB) motion,
the limit-cycle locomotion phase (LLM)
and, for large values of $k\tau$, a region
characterized by chaotic (C) locomotion.
These phases may be characterized by the
standard deviation
$\mathrm{std}(\omega)=\sqrt{\langle\omega^2\rangle-\langle\omega\rangle^2}$
of the angular frequency, which is elevated
in the chaotic and in the FB phase, and small
for LLM. The average angular frequency 
$|\langle\omega\rangle|$
is in contrast highest for limit-cycle locomotion.

Also shown in Fig.~\ref{fig:phaseDiagram} 
is a comparison with the approximate
estimate (\ref{k_critical}) of the transition
between the stationary and the locomotive state,
which is accurate when the transfer function
$y(x)$ can be linearized, viz in the limit of 
a vanishing FB phase. A typical time series of
the motor torque within the chaotic phase
is shown in the right panel of Fig.~\ref{fig:chaotic_attractors}.

\section{Discussion}

Robotic locomotion may be generated either via top-down 
control mechanisms or alternatively via self-organizing 
dynamical processes \citep{aguilar2016review}. In the first, 
more traditional approach, the `brain' of the agent is
responsible for computing the control signals that drive 
the actuators, with error correction occurring through
a high level evaluation of sensory measurements.
The complexity of the control problem can however be 
reduced when robotic behavior is generated via 
self-organizing processes and local instabilities 
of the neural dynamics \citep{der2012playful}.

Our work aims to reduce the theoretical and the 
computational constraints needed to design autonomous 
agents. Based on previous works on barrel- and sphere
shaped robots 
\citep{sandor2015sensorimotor,martin2016closed},
we propose and study a novel actuator for wheeled 
robots. The actuator simulates the physics of the 
transmission rod used by classical steam engines,
being controlled at the same time by only one or two
rate-encoding neurons. Together with the wheel and with
the proprioceptual environmental feedback, the
controlling neurons form an overarching dynamical 
system that generates motion primitives in terms
of stable attractors. The such produced self-organized 
fixpoints, limit cycles and chaotic attractors, correspond 
to non-moving robots, to robots moving with constant 
speed and, respectively, to robots engaging in 
exploratory behaviors. All results are robust to 
noise present either in the environment or in 
the proprioceptual input stream \citep{martin2016closed}.   

A particular feature of the controller proposed here
is that the direction of the movement, forward or 
backward, is selected by breaking time reversal
symmetry. Studying two-wheeled robots we demonstrate 
that attractors corresponding to forward and backward 
motions coexist in the sensorimotor loop, allowing the
robot to change direction autonomously
when colliding with a wall. Switching between stable 
attractors can be achieved furthermore via a higher-order 
top-down control. Implementing frozen and quenched
control signals, we are able to kick the robot
reliably between distinct attractors, that is to
kick the robot from one motor primitive into another
motion primitive. 

Examining in addition the bifurcation diagram leading 
from stationary states characterized by fixpoints 
in the phase space of the sensorimotor loop 
to limit-cycle locomotion, we find that two routes 
to locomotion exist: a single step process
via a saddle node bifurcation of limit cycles,
and a two-step scenario via a supercritical Hopf
bifurcation followed by limit-cycle merging through
distinct homoclinic bifurcations.

The here presented kick-control schemes demonstrate that 
simple impulse-like control signals are sufficient for 
creating complex behaviors whenever the motor primitives 
are given in terms of stable sensorimotor attractors. Hence
one may speculate that kick control could also be used 
effectively in case of more complex robotic architectures, 
possibly in a combination with other types of control schemes
(e.\ g.\ with the KA models \citep{harter2005chaotic}).

The concept of kick control also allows for a more general 
framework which is not necessarily neuro related. 
However, when it comes to the scalability of the proposed 
control scheme to robots with several dozens or even hundreds 
of degrees of freedoms and sensory channels, 
the most convenient underlying dynamical systems 
for the internal controller are adaptive neural 
networks with learning. The attractors there are 
then generalizations of the simple limit-cycle 
and chaotic attractors presented here, but the 
main idea of kicking the phase point from the 
basin of attraction of one attractor to another 
attraction domain remains the same.

Finally, we believe that the proposed model
for generating attractors for locomoting
robots and controlling their motion by kicking
the phase point to their respective basins of
attraction may also be used for teaching dynamical
systems in advanced high school physics courses.
The Lego robots allow for interactive
demonstrations, e. g. in lab activities,
of how attractors may be used in real-world
applications, hence providing an intuitive
understanding of the terminology and underlying
phenomena.

\section*{Conflict of Interest Statement}

The authors declare that the research was 
conducted in the absence of any commercial 
or financial relationships that could be 
construed as a potential conflict of interest.

\section*{Author Contributions}

The manuscript was written by BS and CG.
The experiments were conceived and designed
by CG, BS, TK, MN and LM. Lego experiments
performed mainly by TK and MN; simulations
performed by LM. The data was analyzed by
BS and CG with calculations proposed by MN,
and plots produced by BS and TK.

\section*{Acknowledgments}

The support of the German Science Foundations (DFG)
and discussions with Komal Bhattacharyya are acknowledged.
The work of BS was funded by the Content Pedagogy 
Research Program of the Hungarian Academy of Sciences.

\bibliographystyle{frontiersinSCNS_ENG_HUMS}
\bibliography{wheeled_robots}

\begin{thebibliography}{26}
\providecommand{\natexlab}[1]{#1}
\expandafter\ifx\csname urlstyle\endcsname\relax
  \providecommand{\doi}[1]{doi:\discretionary{}{}{}#1}\else
  \providecommand{\doi}{doi:\discretionary{}{}{}\begingroup
  \urlstyle{rm}\Url}\fi
\providecommand{\selectlanguage}[1]{\relax}
\providecommand{\bibAnnoteFile}[1]{%
  \IfFileExists{#1}{\begin{quotation}\noindent\textsc{Key:} #1\\
  \textsc{Annotation:}\ \input{#1}\end{quotation}}{}}
\providecommand{\bibAnnote}[2]{%
  \begin{quotation}\noindent\textsc{Key:} #1\\
  \textsc{Annotation:}\ #2\end{quotation}}

\bibitem[{Aguilar et~al.(2016)Aguilar, Zhang, Qian, Kingsbury, McInroe,
  Mazouchova et~al.}]{aguilar2016review}
Aguilar, J., Zhang, T., Qian, F., Kingsbury, M., McInroe, B., Mazouchova, N.,
  et~al. (2016).
\newblock {A review on locomotion robophysics: the study of movement at the
  intersection of robotics, soft matter and dynamical systems}.
\newblock \emph{Reports on Progress in Physics} 79, 110001
\bibAnnoteFile{aguilar2016review}

\bibitem[{Conradt et~al.(2015)Conradt, Galluppi, and
  Stewart}]{conradt2015trainable}
Conradt, J., Galluppi, F., and Stewart, T.~C. (2015).
\newblock Trainable sensorimotor mapping in a neuromorphic robot.
\newblock \emph{Robotics and Autonomous Systems} 71, 60--68
\bibAnnoteFile{conradt2015trainable}

\bibitem[{Das et~al.(2006)Das, Kar, and Chaudhury}]{das2006simple}
Das, T., Kar, I., and Chaudhury, S. (2006).
\newblock Simple neuron-based adaptive controller for a nonholonomic mobile
  robot including actuator dynamics.
\newblock \emph{Neurocomputing} 69, 2140--2151
\bibAnnoteFile{das2006simple}

\bibitem[{Der and Martius(2012)}]{der2012playful}
Der, R. and Martius, G. (2012).
\newblock \emph{The Playful Machine: Theoretical Foundation and Practical
  Realization of Self-Organizing Robots}, vol.~15 (Springer Science \& Business
  Media)
\bibAnnoteFile{der2012playful}

\bibitem[{Ghazi-Zahedi et~al.(2017)Ghazi-Zahedi, Langer, and
  Ay}]{ghazi2017morphological}
Ghazi-Zahedi, K., Langer, C., and Ay, N. (2017).
\newblock Morphological computation: Synergy of body and brain.
\newblock \emph{Entropy} 19, 456
\bibAnnoteFile{ghazi2017morphological}

\bibitem[{Gros(2014)}]{gros2014generating}
Gros, C. (2014).
\newblock Generating functionals for guided self-organization.
\newblock In \emph{Guided Self-Organization: Inception}, ed. M.~Prokopenko
  (Springer). 53--66
\bibAnnoteFile{gros2014generating}

\bibitem[{Gros(2015)}]{gros2015complex}
Gros, C. (2015).
\newblock \emph{Complex and adaptive dynamical systems: A primer} (Springer)
\bibAnnoteFile{gros2015complex}

\bibitem[{Harter and Kozma(2005)}]{harter2005chaotic}
Harter, D. and Kozma, R. (2005).
\newblock {Chaotic neurodynamics for autonomous agents}.
\newblock \emph{IEEE Transactions on Neural Networks} 16, 565--579
\bibAnnoteFile{harter2005chaotic}

\bibitem[{Hennig(2013)}]{hennig2013theoretical}
Hennig, M.~H. (2013).
\newblock Theoretical models of synaptic short term plasticity.
\newblock \emph{Frontiers in computational neuroscience} 7
\bibAnnoteFile{hennig2013theoretical}

\bibitem[{Ijspeert et~al.(2002)Ijspeert, Nakanishi, and
  Schaal}]{ijspeert2002learning}
Ijspeert, A.~J., Nakanishi, J., and Schaal, S. (2002).
\newblock {Learning Attractor Landscapes for Learning Motor Primitives}.
\newblock In \emph{Advances in Neural Information Processing Systems} (MIT
  Press), 1547--1554
\bibAnnoteFile{ijspeert2002learning}

\bibitem[{Jimenez-Fernandez et~al.(2012)Jimenez-Fernandez, Jimenez-Moreno,
  Linares-Barranco, Dominguez-Morales, Paz-Vicente, and
  Civit-Balcells}]{jimenez2012neuro}
Jimenez-Fernandez, A., Jimenez-Moreno, G., Linares-Barranco, A.,
  Dominguez-Morales, M.~J., Paz-Vicente, R., and Civit-Balcells, A. (2012).
\newblock {A neuro-inspired spike-based PID motor controller for multi-motor
  robots with low cost FPGAs}.
\newblock \emph{Sensors} 12, 3831--3856
\bibAnnoteFile{jimenez2012neuro}

\bibitem[{Koz{\l}owski and Pazderski(2004)}]{kozlowski2004modeling}
Koz{\l}owski, K. and Pazderski, D. (2004).
\newblock Modeling and control of a 4-wheel skid-steering mobile robot.
\newblock \emph{International journal of applied mathematics and computer
  science} 14, 477--496
\bibAnnoteFile{kozlowski2004modeling}

\bibitem[{Kubisch et~al.(2011)Kubisch, Werner, and Hild}]{kubisch2011using}
Kubisch, M., Werner, B., and Hild, M. (2011).
\newblock Using co-existing attractors of a sensorimotor loop for the motion
  control of a humanoid robot.
\newblock In \emph{IJCCI (NCTA)}. 385--390
\bibAnnoteFile{kubisch2011using}

\bibitem[{Linkerhand and Gros(2013)}]{linkerhand2013generating}
Linkerhand, M. and Gros, C. (2013).
\newblock Generating functionals for autonomous latching dynamics in attractor
  relict networks.
\newblock \emph{Scientific reports} 3
\bibAnnoteFile{linkerhand2013generating}

\bibitem[{Mandow et~al.(2007)Mandow, Martinez, Morales, Blanco, Garcia-Cerezo,
  and Gonzalez}]{mandow2007experimental}
Mandow, A., Martinez, J.~L., Morales, J., Blanco, J.~L., Garcia-Cerezo, A., and
  Gonzalez, J. (2007).
\newblock Experimental kinematics for wheeled skid-steer mobile robots.
\newblock In \emph{IEEE/RSJ International Conference on Intelligent Robots and
  Systems} (IEEE), 1222--1227
\bibAnnoteFile{mandow2007experimental}

\bibitem[{Martin et~al.(2016)Martin, S{\'{a}}ndor, and Gros}]{martin2016closed}
Martin, L., S{\'{a}}ndor, B., and Gros, C. (2016).
\newblock {Closed-loop robots driven by short-term synaptic plasticity:
  Emergent explorative vs. limit-cycle locomotion}.
\newblock \emph{Frontiers in Neurorobotics} 10, 12
\bibAnnoteFile{martin2016closed}

\bibitem[{Martius et~al.(2013)Martius, Der, and Ay}]{martius2013information}
Martius, G., Der, R., and Ay, N. (2013).
\newblock Information driven self-organization of complex robotic behaviors.
\newblock \emph{PLOS ONE} 8, e63400
\bibAnnoteFile{martius2013information}

\bibitem[{Pfeifer et~al.(2007)Pfeifer, Lungarella, and Iida}]{pfeifer2007self}
Pfeifer, R., Lungarella, M., and Iida, F. (2007).
\newblock Self-organization, embodiment, and biologically inspired robotics.
\newblock \emph{science} 318, 1088--1093
\bibAnnoteFile{pfeifer2007self}

\bibitem[{Pisarchik and Feudel(2014)}]{pisarchik2014control}
Pisarchik, A.~N. and Feudel, U. (2014).
\newblock {Control of multistability}.
\newblock \emph{Physics Reports} 540, 167--218
\bibAnnoteFile{pisarchik2014control}

\bibitem[{Prokopenko et~al.(2009)Prokopenko, Boschetti, and
  Ryan}]{prokopenko2009information}
Prokopenko, M., Boschetti, F., and Ryan, A.~J. (2009).
\newblock An information-theoretic primer on complexity, self-organization, and
  emergence.
\newblock \emph{Complexity} 15, 11--28
\bibAnnoteFile{prokopenko2009information}

\bibitem[{S{\'a}ndor and Gros(2015)}]{sandor2015versatile}
S{\'a}ndor, B. and Gros, C. (2015).
\newblock A versatile class of prototype dynamical systems for complex
  bifurcation cascades of limit cycles.
\newblock \emph{Scientific reports} 5, 12316
\bibAnnoteFile{sandor2015versatile}

\bibitem[{S{\'a}ndor et~al.(2015)S{\'a}ndor, Jahn, Martin, and
  Gros}]{sandor2015sensorimotor}
S{\'a}ndor, B., Jahn, T., Martin, L., and Gros, C. (2015).
\newblock The sensorimotor loop as a dynamical system: How regular motion
  primitives may emerge from self-organized limit cycles.
\newblock \emph{Frontiers in Robotics and AI} 2, 31
\bibAnnoteFile{sandor2015sensorimotor}

\bibitem[{Tani and Ito(2003)}]{tani2003self}
Tani, J. and Ito, M. (2003).
\newblock Self-organization of behavioral primitives as multiple attractor
  dynamics: A robot experiment.
\newblock \emph{IEEE Transactions on Systems, Man, and Cybernetics-Part A:
  Systems and Humans} 33, 481--488
\bibAnnoteFile{tani2003self}

\bibitem[{Todorov(2004)}]{todorov2004optimality}
Todorov, E. (2004).
\newblock Optimality principles in sensorimotor control.
\newblock \emph{Nature neuroscience} 7, 907--915
\bibAnnoteFile{todorov2004optimality}

\bibitem[{Toutounji and Pasemann(2014)}]{toutounji2014behavior}
Toutounji, H. and Pasemann, F. (2014).
\newblock Behavior control in the sensorimotor loop with short-term synaptic
  dynamics induced by self-regulating neurons.
\newblock \emph{Frontiers in neurorobotics} 8
\bibAnnoteFile{toutounji2014behavior}

\bibitem[{Voicu et~al.(2004)Voicu, Kozma, Wong, and Freeman}]{voicu2004spatial}
Voicu, H., Kozma, R., Wong, D., and Freeman, W.~J. (2004).
\newblock {Spatial navigation model based on chaotic attractor networks}.
\newblock \emph{Connection Science} 16, 1--19
\bibAnnoteFile{voicu2004spatial}

\end{thebibliography}

\end{document}